\documentclass[%
aip,jcp,amsmath,amssymb, reprint
]{revtex4-2}
\usepackage{graphicx}
\usepackage{dcolumn}
\usepackage{bm}
\usepackage{amssymb}
\usepackage{amsmath}
\usepackage{supertabular}
\usepackage[table,x11names]{xcolor}
\usepackage{longtable}
\usepackage{colortbl}
\usepackage{multirow}\usepackage{braket}
\usepackage{esvect}
\usepackage{gensymb}
\usepackage{soul}
\let\oldAA\AA
\renewcommand{\AA}{\text{\normalfont\oldAA}}

\usepackage{qtree}
\usepackage{tipa}
\usepackage{appendix}
\usepackage{ulem}
\usepackage{xcolor}
\usepackage{tabularx,booktabs}
\usepackage{simplewick}
\usepackage{simpler-wick}
\usepackage{multirow, makecell}
\usepackage{scalerel}
\usepackage[framemethod=tikz]{mdframed}

\usepackage[colorinlistoftodos]{todonotes}
\presetkeys{todonotes}{inline,backgroundcolor=white}{}
\begin{document}
\title{Noise-independent Route towards the Genesis of a COMPACT Ansatz for Molecular Energetics: a Dynamic Approach}
\author{Dipanjali Halder$^{1}$, Dibyendu Mondal$^{1}$, Rahul Maitra$^{1,2,\dagger}$}
\affiliation{$^1$ Department of Chemistry, \\ Indian Institute of Technology Bombay, Powai, Mumbai 400076, India \\
$^2$ Centre of Excellence in Quantum Information, Computing, Science and Technology, \\Indian Institute of Technology Bombay, Powai, Mumbai 400076, India \\
$^\dagger$ rmaitra@chem.iitb.ac.in}


\begin {abstract}
Recent advances in quantum information and quantum science have inspired the development of various compact dynamic structured ans\"{a}tze that are expected to be realizable in the Noisy Intermediate-Scale Quantum (NISQ) devices. However, such ans\"{a}tze construction strategies hitherto developed involve considerable measurements, and thus they deviate significantly in NISQ platform from their ideal structures. Therefore, it is imperative that the usage of quantum resources must be minimized while retaining the expressivity and dynamical structure of the ansatz that can adapt itself depending on the degree of correlation. We propose a novel ansatz construction strategy based on the \textit{ab-initio} many-body perturbation theory that requires \textit{no} pre-circuit measurement and thus it remains structurally unaffected by any hardware noise. The accuracy and quantum complexity associated with the ansatz are solely dictated by a pre-defined perturbative order as desired and hence are tunable. Furthermore, the underlying perturbative structure of the ansatz construction pipeline enables us to decompose any high-rank excitation that appears in higher perturbative orders into the product of various low-rank operators, and it thus keeps the execution gate-depth to its minimum. With a number of challenging applications on strongly correlated systems, we demonstrate that our ansatz performs significantly better, both in terms of accuracy, parameter count and circuit depth, in comparison to the allied unitary coupled cluster based ans\"{a}tze.    
\end{abstract}

\maketitle
\section{Introduction} 
Quantum computers have surfaced as a promising computational 
platform that possesses enormous potential in tackling 
challenges that are beyond the capabilities of conventional 
classical computers. Simulation of many-body quantum systems 
is particularly challenging for classical computers due to 
the exponential scaling of the computational complexities. 
Hence, many-body problems pertinent to the field of quantum
chemistry, condensed matter physics, and materials science is 
regarded as a crucial near-term application of the quantum 
computers and is believed to be among the initial domains 
to exploit ``quantum advantage"~\cite{RevModPhys.92.015003}.\\
Owing to the recent experimental breakthroughs in the realm 
of quantum technologies, quantum computers have stepped into 
the noisy intermediate-scale quantum (NISQ) era, in which the 
quantum hardwares are unfortunately not yet capable of 
attaining fault-tolerant quantum computations. The 
contemporary state-of-the-art NISQ hardwares are known to 
suffer from short coherence times, limited number of qubits, 
imprecise readouts and noisy operations, which are the major 
obstacles towards the practical demonstration of ``quantum 
advantage". In order to alleviate the limitations of the
imperfect NISQ hardwares, the preceding decade has witnessed 
the emergence of hybrid quantum-classical algorithms, viz. 
variational quantum algorithms (VQAs). 
The first VQA, namely variational 
quantum eigensolver (VQE)~\cite{Peruzzo_2014} was proposed by Peruzzo and co-
workers in 2014 to address the electronic structure problem, 
i.e., to solve for the ground state of the many-electron
Schrodinger equation. The algorithm was first experimentally 
realized on a photonic quantum computer~\cite{Peruzzo_2014}, followed by its 
successful demonstrations on superconducting~\cite{PhysRevX.6.031007} and trapped ion 
quantum hardwares~\cite{PhysRevA.95.020501,PhysRevX.8.031022}. VQE, which is built upon the fundamental 
concept of variational principle can be realized as a 
general framework that aims to find the best variational 
approximation to the ground state energy of a given  
Hamiltonian through an iterative minimization of energy 
expectation value with a trial wavefunction ansatz, encoded 
via a parametrized quantum circuit. VQE relies on both 
quantum resources, quantified by circuit complexities 
associated with the parametrized ansatz and classical 
resources, quantified by the speed at which the minimization 
process converges. 
This distinct hybrid nature of VQE primarily accounts for its 
compatibility with the current pre fault-tolerant NISQ 
hardware. However, the feasibility of its implementation relies on 
the circuit complexities of the ansatz. An ansatz not only
dictates the feasibility of implementation of VQE, it also 
significantly impacts the accuracy of estimated energies 
and optimization landscapes~\cite{TILLY20221}. Hence, it is imperative to
tailor an ansatz that balances the trade-offs between 
accuracies and circuit complexities. Over the last few years, 
there have been varied developments in pursuit of this goal,
with the chemically motivated unitary coupled cluster (UCC)~\cite{Peruzzo_2014,D1CS00932J}
ansatz being the first one to be introduced and studied in 
the framework of VQE. Although UCC theory was originally 
proposed decades back, it recently experienced a massive 
resurgence of interest post its appearance in the seminal 
work of Peruzzo \textit{et al}~\cite{Peruzzo_2014}. 
UCC truncated 
at singles and doubles excitation level (UCCSD) is often 
considered as a ubiquitous choice for compact representation 
of many-electron wavefunction. However, UCCSD turns out to be 
not expressive enough, especially in the regions of molecular 
strong correlation. 
Drawing inspirations from Nakatsuji's density theorem~\cite{PhysRevA.14.41} and 
Nooijen's conjecture~\cite{PhysRevLett.84.2108}, one may improve the expressibility of 
UCCSD framework by replacing the usual rank-one and rank-two
hole-particle excitations by the generalized rank-one and 
generalized rank-two operators, respectively. While this is 
expected to recover most of the 
missing correlation of UCCSD, it also comes with tremendous 
computational overhead. A recent study carried out by Lee 
and co-workers have focused on retaining only paired 
generalized double ``excitations", giving 
rise to the k-UpCCGSD~\cite{doi:10.1021/acs.jctc.8b01004} ansatz for which the circuit depth
theoretically scales linearly. With a similar motivation,
few of the present authors have introduced a unitarized 
dual exponential coupled cluster based ansatz, viz., 
UiCCSDn~\cite{10.1063/5.0114688}, which is capable of 
simulating the high-rank correlation effects employing rank-
one and rank-two parametrization via nested commutators 
between usual rank-two cluster operators and rank-two vacuum-
annihilating scattering operators. Aiming towards further 
compactification of the ansatz for NISQ realization, 
Grimsley \textit{et. al.}~\cite{Grimsley_2019} introduced a dynamically structured 
ansatz, namely ADAPT-VQE that iteratively selects the ``best" 
set of operators through energy gradient evaluation. 
Although, ADAPT-VQE is capable of extracting the correlation 
energy employing a very small subset of operators, the 
operator selection procedure itself incurs a very high 
measurement overhead. Following the advent of ADAPT-VQE, 
there has been numerous studies focusing on the ways to 
minimize the measurement overhead associated with the 
operator selection protocol. To this end, the present authors 
have proposed a dynamic ansatz construction recipe, COMPASS~\cite{10.1063/5.0153182},
which employs the energy sorting and operator commutativity 
prescreening criteria to span the $N$-electron Hilbert space
at bare minimum circuit complexities and measurement 
overhead. In an attempt to minimize the quantum complexities associated with VQE, Metcalf \textit{et. al.} introduced the double unitary coupled cluster (DUCC)~\cite{doi:10.1021/acs.jctc.0c00421, Bauman2022} method to downfold the correlation effects into an active space of reduced dimensionality, leading to a resource efficient approach. In 2022, Fedorov and co-workers came up with the 
unitary selective coupled cluster (USCC)~\cite{Fedorov2022unitaryselective} approach inspired by 
the fundamental ideas of selective heat-bath configuration 
interaction (SHCI). USCC employs the electronic Hamiltonian 
matrix elements and amplitudes for excitation operators to 
determine the ``important" excitations to tailor the optimal 
ansatz. Very recently, Haider \textit{et. al.}~\cite{doi:10.1021/acs.jpca.3c01753} introduced 
compactification of trotterized UCCSDT ansatz via restricting 
the operators by spin and orbital symmetries. 
Fan \textit{et. al.}~\cite{fan2023circuitdepth} proposed an energy sorting procedure 
for operator screening within the UCC framework towards the 
development of a compact and resource efficient ansatz.\\

The various static structured ans\"{a}tze like UCCSD or UCCSDT include
several operators which are not significant or physically meaningful; 
however, their inclusion 
leads to a pronounced proliferation of gate depth. Thus NISQ devices prefer
expressive yet compact ansatz with minimal number of relevant parameters that may be
selected dynamically depending on the system under consideration. However, most of the dynamic methods hitherto developed 
require substantial pre-circuit measurements to construct 
the unitary which add significantly to computation overhead. 
More importantly, such a dynamic construction of the ansatz 
(ADAPT-VQE or COMPASS) was proposed based on an ideal quantum 
environment. However, in the present noisy quantum 
architecture, a 
measurement based dynamic circuit tailoring protocol may 
need to undergo a paradigm shift in the way the operators are 
selected from the pool as the unitary generated in NISQ hardware 
is often far from being optimal. Thus, one must minimize the usage 
of quantum resources for such dynamic ansatz construction,
and instead, should rely more on first-principle based or 
intuitive approaches that may navigate past any potential
pitfall posed by the NISQ architecture~\cite{D3SC05807G}. This paves the motivation to 
this development: we, for the first time, develop a first-principle
based strategy starting from many-body perturbation theory (MBPT)~\cite{lindgren1986first, shavitt_bartlett_2009} 
that dynamically constructs an extremely compact yet highly 
expressive ansatz \textit{without any pre-circuit
measurement.} At this stage, one should note that it is the \textit{ansatz construction pipeline} that is 
measurement-free; however, the iterative optimization of the ansatz 
parameters would still require measurements. 
We term such an ansatz construction protocol as
Cognitive Optimization via Many-body Perturbation theory towards Ansatz 
Construction and Tailoring (COMPACT). \\

In the following section, we first briefly discuss the
preliminaries of VQE which also concisely introduces the approach we
are going to adopt. The generation of the 
COMPACT dynamic ansatz pivoted upon the MBPT will subsequently 
be expounded in an order-by-order manner in Sec. \ref{opselect}. 
In Section \ref{res}, we will
discuss the general implementation strategy, followed by an in-depth 
analysis of the performance of COMPACT in determining the 
ground state energies of a few strongly correlated molecules. Finally, in 
Section \ref{concl}, we would summarize our findings along with introducing
a roadmap towards possible further developments.

\section{Theory}
\subsection{Preliminaries on Variational Quantum Eigensolver (VQE) algorithm and the motivation towards the development of MBPT guided ansatz:}
Within this section, we aim to explicate the key fundamentals of the VQE algorithm. We begin by revisiting what lies at the very core of VQE i.e., the variational principle, which states that the exact ground state energy of a system is the global minima of the expectation value of the system Hamiltonian. In mathematical terms, variational principle can be expressed as follows:
\begin{equation}
    E_{0} \leq \underset{\Phi}{min} \bra{\Phi}H\ket{\Phi}
\end{equation}
Here $E_{0}$ denotes the true ground state energy of a 
molecular system of interest, $\Phi$ represents a normalized 
trial wave function and $H$ denotes the electronic
Hamiltonian of the concerned molecular system, whose representation in the language of second quantized operators is given as: 
\begin{eqnarray}
H=\sum\limits_{pq}h_{pq}a_{p}^{\dagger}a_{q}+\frac{1}{4}\sum\limits_{pqrs}h_{pqrs}a_{p}^{\dagger}a_{q}^{\dagger}a_{s}a_{r}.
\end{eqnarray}
where $h_{pq}$ and $h_{pqrs}$ correspond to one- and two-electron integrals respectively, with \textit{p,q,r,s} indicating general spinorbital indices.\\
VQE leverages the fundamental concept of the variational principle to obtain an upper bound estimate for the elusive ground state energy of a many-electron system. Its essence lies in the meticulous selection of the initial ``guess" for the target ground state wavefunction, commonly referred to as the trial wavefunction ansatz. On a quantum hardware, an ansatz is realized via a parametrized quantum circuit, encompassing a series of parametrized unitary operations. In order to approximate the true ground state energy, VQE intends to search for the optimal set of parameters, i.e., the parameters which lead to minimization of the expectation value of the Hamiltonian. This is achieved via a classical optimizer, which adjusts the parameters of the ansatz iteratively, till the energy expectation value converges to a minimum. A succinct mathematical representation of the overall VQE algorithm is given by 
\begin{eqnarray}
    E_{0}=\underset{\vec{\theta}}{min}\bra{\Phi_{HF}}U^{\dagger}(\vec{\theta})HU(\vec{\theta})\ket{\Phi_{HF}}
\label{cost_function}
\end{eqnarray}
where, $\ket{\Phi_{HF}}$ refers to the Hartree-Fock reference determinant which is commonly used to represent the closed shell many-electron wavefunction, $U(\vec{\theta})$ refers to the series of parameterized unitary operations and  $\vec{\theta}$ refers to the set of variational parameters.\\ 
A salient feature of VQE is its hybrid nature, making it
amenable to the near term noisy quantum devices, albeit the
degree of feasibility depends on the choice of the ansatz 
to a great extent. Expectedly, a shallow depth ansatz would
enhance the amenability for NISQ realization and vice versa. 
Hence, an ansatz is considered to be the prime ingredient 
toward the successful implementation of VQE. A succinct 
overview of all the noteworthy ans\"{a}tze in the context of 
VQE have been well documented in ~\cite{D1CS00932J, TILLY20221}. In
this section we restrict ourselves to a fundamental chemistry
inspired unitary coupled cluster ansatz (UCC), as it acts 
as the cornerstone towards COMPACT.  \\

The UCC ansatz employs an exponential operator, where the 
exponent is taken to be an anti-hermitian sum of 
hole-particle excitation and de-excitation operators 
\begin{eqnarray}
    \ket{\Phi_{UCC}} &=& e^{T(\theta)-T^{\dagger}(\theta)}\ket{\Phi_{HF}}\\
    &=& e^{\tau(\theta)}\ket{\Phi_{HF}}
    \label{UCC}
\end{eqnarray}
where the cluster operator $T$ is defined as: $T=T_{1}+T_{2}+T_{3}+\hdots$.   
The excitation operator $T_{n}$ generates the $n-$th excited determinants from the Hartree-Fock reference.
\begin{eqnarray}
    T_{1}\ket{\Phi_{HF}} &=& \sum\limits_{i;a}\theta_{i}^{a}\ket{\Phi_{i}^{a}}\\
    T_{2}\ket{\Phi_{HF}} &=& \sum\limits_{i,j;a,b}\theta_{ij}^{ab}\ket{\Phi_{ij}^{ab}}\\
    \nonumber
    &\vdots&
\end{eqnarray}
where the indices $i,j$ refer to occupied spinorbitals and 
$a,b$ refer to unoccupied spinorbitals with respect to the 
Hartree Fock reference. The amplitudes $\theta_{i}^{a}$ and 
$\theta_{ij}^{ab}$ refer to the cluster amplitudes which act 
as the variational parameters. It is a common practice to 
restrict the rank of $T$ to be two-body resulting in UCCSD.
UCCSD is known to have limited expressibility, thereby suffers
from restricted applicability for strong correlation.

One of the first approaches towards the 
compactification of the UCCSD ansatz was proposed by Romero~\cite{Romero_2019} who restricted the number of cluster operators that 
enters in Trotterized UCCSD by screening them through their 
first order perturbative values. The authors had demonstrated 
that such a restriction on the parameters results in drastic
compactification of the ansatz without unduly sacrificing
the characteristic accuracy of the parent UCCSD. One may 
think of its higher order generalizations and selectively 
include (explicitly or implicitly) the \textit{important} 
rank-three, four,... cluster operators by screening 
them from their respective leading perturbative order. Our
endeavor is a step towards such an intuitive construction 
of the dynamic ansatz which is intimately guided by 
many-body perturbation theory. Furthermore, we will 
decompose any such high rank excitation terms through
commutators of certain low rank operators in an optimal
way such that the circuit depth remains at its minimum.
Such a dynamic construction requires \textit{no} quantum 
measurements, safeguarding it against any uncertainty that
a measurement-based construction of a
dynamic ansatz ought to suffer in the NISQ devices.

\begin{figure*}
    \includegraphics[width=17cm, height=15.25cm]{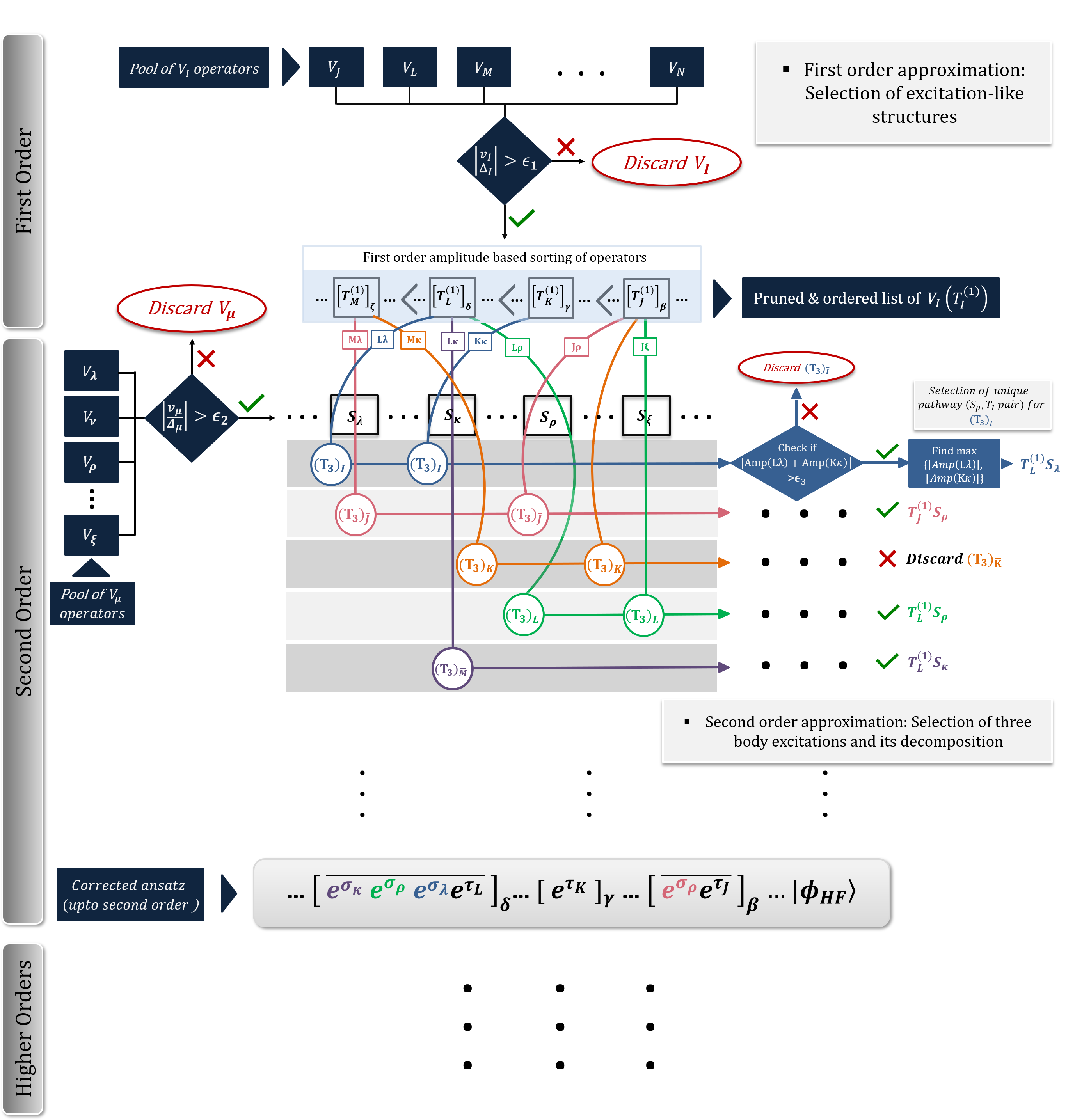}
    \caption{\textbf{The hierarchical structure of the COMPACT ansatz construction pipeline at various perturbative orders. The curved lines connecting the various $T_I$'s (generated at the first order) to some selected $S_\mu$'s are based on the commonality of the contractible indices and such a path leading to the generation of triples are generically denoted as $(I\mu)$. For each $T_3$, all such contributions from various paths are summed over to check if the resultant is greater than a threshold $\epsilon_3$ and only the most dominant contributing path is retained. The generation of singles is not shown for brevity. The procedure is repeated for generating higher order terms.}}
    \label{fig:algorithm}
\end{figure*}

\subsection{Operators selection and ansatz construction pipeline towards COMPACT:}
\label{opselect}
For UCCSD ansatz, a common practice is to initialize the
doubles amplitudes from their first order perturbed values 
that contribute to the MP2 energy. In most of the closed 
shell applications, the Fock operator is often chosen as 
the unperturbed Hamiltonian while the perturbation $V$ 
is taken to be the difference between the electron-electron
Coulomb interaction and the Hartree-Fock potential.
\begin{eqnarray}    
V=\sum\limits_{i}^{N}\sum\limits_{j>i}^{N}\frac{1}{|\vec{r_{i}}-\vec{r_{j}}|}-\sum\limits_{i}^{N}v_{HF}(i)
\end{eqnarray}
A perturbative expansion of the wavefunction in terms of
$V$ thus gives us a recipe to decipher which set of 
operators first appear at various perturbative orders. This
leads one to suitably approximate and filter various rank
of excitations based on the respective lowest orders in which 
they appear to tailor an ansatz. In the following,
motivated by MBPT, we take a bottom-up approach where 
we hierarchically move upwards the perturbation ladder
to select the most important excitations and dynamically
construct the ansatz. 

\subsubsection{First-order perturbative correction to the wavefunction: Selection of the double excitation manifold}

The selection of the two-body operators 
based on their first order estimates is well studied
and has been previously employed to reduce the parameter 
space\cite{Romero_2019}. While this is well-understood and obvious, we still 
briefly discuss this formally from the perturbation theoretic
standpoint such that we maintain continuity in the next
section.

With the choice of the Hartree-Fock determinant $\ket{\Phi_{HF}}$
taken to be the unperturbed function $\ket{\Psi_{0}^{(0)}}$, 
the first order correction to the ground state wavefunction
may be expressed in terms of $\ket{\Psi_{n}^{(0)}}$, 
the zeroth order eigenfunctions of the Fock operator:
\begin{eqnarray}
    \ket{\Psi_{0}^{(1)}} &=& \sum\limits_{n\neq 0}\left[\frac{\bra{\Psi_{n}^{(0)}}V\ket{\Psi_{0}^{(0)}}}{E_{0}^{(0)}-E_{n}^{(0)}}\right]\ket{\Psi_{n}^{(0)}} \\&=& \sum\limits_{n\neq 0}\left[\frac{V_{n0}}{{\Delta E_{0n}^{(0)}}}\right]\ket{\Psi_{n}^{(0)}}
    \label{first_order}
\end{eqnarray}
One may note that $\ket{\Psi_{n}^{(0)}}$ are orthogonal to $\ket{\Psi_{0}^{(0)}}$. 
Since the perturbation $V$ is two-body,
the underlying Slater rule ensures that $\ket{\Psi_{n}^{(0)}}$s
are only the determinants that span the double excitation
manifold. Without further elaboration, Eq. \ref{first_order}
may be written in terms of the many-body basis as:
\begin{eqnarray}
    \ket{\Psi_{0}^{(1)}} = \frac{1}{4}\sum\limits_{ijab}\frac{v_{ij}^{ab}}{\Delta_{ij}^{ab}}\ket{\Phi_{ij}^{ab}} = \frac{1}{4}\sum\limits_{I}\frac{v_{I}}{\Delta_{I}}\ket{\Phi_{I}}
\end{eqnarray}
Here the index $I$ is the four orbital index tuple $({abij})$.

Since the double excitations $T_2$ appear first in the leading perturbative order,
choosing them at the very first step of our ansatz construction pipeline is 
logical. However, instead of an impetuous incorporation of the complete pool 
of $T_{2}$ operators, we assess the importance or the effectiveness of each 
of the $T_{2}$ operators to ensure a shallow depth ansatz: only those double 
excitation operators are retained whose corresponding absolute value of amplitude
at the first order perturbative estimate, $t_I^{(1)}=|v_I/\Delta_I| > \epsilon_{1}$, 
where $\epsilon_1$ is a suitable user-defined threshold and the rest are discarded. 
With $N_D$ number of 
such operators being screened, we align them in descending order of their absolute 
values of first order perturbative estimates and place them one by 
one in an ordered set of operator blocks. These operator ``blocks" may be defined as
snippets of the final ansatz, each of which starts with one cluster operator 
selected at the first order. In what follows, some of these individual blocks 
will be subsequently expanded via 
second and higher order screening to induce specific higher rank excitations when 
acted upon the reference function. Towards this, to begin with 
$N_D$ cluster operators with $...|t_J^{(1)}| > ... > |t_K^{(1)}|> ... > |t_L^{(1)}|$, 
$e^{\tau_{J}}$ is placed in an operator block $\beta$,..., 
$e^{\tau_{K}}$ in operator block $\gamma$, ... and so on 
in a disentangled manner with the operator blocks 
arranged as $...\delta > ...>\gamma>...>\beta>...>1$. 
More explicitly, the operator strings in operator block
$\beta$ act before those in $\gamma$ and so on.
Thus the unitary constructed through the inclusion of the selected 
operators at the first MBPT order may be expressed as:
\begin{eqnarray}
    U^{[1]} &=& ...\Big[e^{\tau_L}\Big]_{\delta}... \Big[e^{\tau_K}\Big]_{\gamma}... \Big[e^{\tau_J}\Big]_{\beta}...\nonumber \\
    U^{[1]} &=& \prod_{\alpha=1}^{N_D} \Big[e^{\tau_I}\Big]_{\alpha}
    \label{unitary1}
\end{eqnarray}
Here the indices $J, K, L$ are the four orbital index tuple 
associated with the selected $T_2$ operators and 
$\tau=T-T^\dagger=\theta\hat{\kappa}$ is the anti-hermitian
operator where $\hat{\kappa}$ is the operator 
string and $\theta$ is the parameter. One may note that such 
a restriction on the magnitude of the first order estimate
of the two-body cluster amplitudes screens out any 
symmetry-disallowed excitations and thus no additional
pruning needs to be performed on top.

\subsubsection{Second-order perturbative correction to the wavefunction: Selection of the single and triple excitation manifolds}

The second order correction to the ground state wavefunction is given by:
\begin{eqnarray}
    \ket{\Psi_{0}^{(2)}} &=& \sum\limits_{n\neq 0}\Bigg[\sum\limits_{m\neq 0}\frac{\bra{\Psi_{n}^{(0)}}V\ket{\Psi_{m}^{(0)}}\bra{\Psi_{m}^{(0)}}V\ket{\Psi_{0}^{(0)}}}{(E_{0}^{(0)}-E_{n}^{(0)})(E_{0}^{(0)}-E_{m}^{(0)})}\nonumber
    \\&-&\frac{\bra{\Psi_{0}^{(0)}}V\ket{\Psi_{0}^{(0)}}\bra{\Psi_{n}^{(0)}}V\ket{\Psi_{0}^{(0)}}}{(E_{0}^{(0)}-E_{n}^{(0)})^{2}} \Bigg]\ket{\Psi_{n}^{(0)}}\\
    &=& \sum\limits_{n\neq 0}\Bigg[\sum\limits_{m\neq 0}\frac{V_{nm}V_{m0}}{\Delta E_{0n}^{(0)}\Delta E_{0m}^{(0)}}-\frac{V_{00}V_{n0}}{(\Delta E_{0n}^{(0)})^{2}} \Bigg]\ket{\Psi_{n}^{(0)}}\nonumber\\
    \label{second_order}
\end{eqnarray}
Clearly, with $V$ taken to be the two-body operator, $\ket{\Psi_{m}^{(0)}}$ can span 
the orthogonal doubly excited space while $\ket{\Psi_{n}^{(0)}}$ are the determinants
that span single, double and triple excitation manifold. Note that the doubles are 
already selected in our ansatz through first order MBPT screening and no more
alteration to the two-body excitation operator pool will be done. We will not account for the second term of Eq. \ref{second_order} as such unlinked terms do not explicitly appear at the second order when the Hamiltonian 
is taken to be normal ordered with respect to the Hartree-Fock reference~\cite{shavitt_bartlett_2009}.
From this second order analysis, we concentrate to select only the dominant single 
and triple excitation operators which appear for the first time in the perturbative expansion.

Taking $\ket{\Psi_{n}^{(0)}}$ to be the triply excited determinants, we first prune
the three-body excitation manifold depending on the orbital indices. 
The determinant $\ket{\Psi_{n}^{(0)}}$ is
generated by the \textit{connected} action of the perturbation operator $V$ on the doubly excited 
determinants $\ket{\Psi_{m}^{(0)}}$; thus there must be one
orbital label (either hole or particle) common between the two
perturbation operators $V$ (one that connects
$\ket{\Psi_{m}^{(0)}}$ and $\ket{\Psi_{0}^{(0)}}$ and the other 
that connects $\ket{\Psi_{m}^{(0)}}$ and 
$\ket{\Psi_{n}^{(0)}}$). This common 
orbital label may be considered to be a contractible orbital index. In what follows, 
among the six orbital indices (3 holes and 3 particles) that characterize each of the triply 
excited determinants, a set of three orbital tuple 
(either \textit{2h1p} or \textit{2p1h}) comes from
each $V$. 
An expansion of Eq. \eqref{second_order} in the many-body basis would make this evident. 
\begin{eqnarray}
    \ket{\Psi_{0}^{(2)}} \xleftarrow{} \sum\limits_{ijkabc}\Bigg[\sum\limits_{m}\frac{v_{ij}^{am}v_{mk}^{bc}}{\Delta_{ijk}^{abc} \Delta_{mk}^{bc}} +\sum\limits_{e}\frac{v_{ie}^{ab}v_{jk}^{ec}}{\Delta_{ijk}^{abc} \Delta_{jk}^{ec}}\Bigg]\ket{\Phi_{ijk}^{abc}}
    \label{second_order_wf}
    \nonumber
    \\
\end{eqnarray}    
which together can be compactly written as:
\begin{eqnarray}
    T_{\overline{K}} \leftarrow{} \frac{(\contraction{}{V_{\mu}}{}{V_I} V_\mu V_I)_{\overline{K}}}{\Delta_{\overline{K}} \Delta_I}= \frac{(\contraction{}{V_\mu}{}{T_I^{(1)}} V_\mu T_I^{(1)})_{\overline{K}}}{\Delta_{\overline{K}}}
    \label{T_3_compact}
\end{eqnarray}
where $\overline{K}$ stands for the six orbital index tuple
associated with the triple excitation, and $\mu, I$
have similar meaning for two-body operators of various structures. The 
overbarred capitalized indices will generically denote three body excitation structure. 
Note that the four orbital tuple that characterize the $V$ which connects $\ket{\Psi_{m}^{(0)}}$ and 
$\ket{\Psi_{n}^{(0)}}$ is denoted by Greek letter $\mu$ to distinguish it from an excitation-like 
structure of $V_I$. To select and parametrize the 
corresponding triple excitations, we use the following 
screening criteria:
\begin{itemize}
    \item \textbf{\textit{Orbital index based excitation subspace pruning:}} With the selected $N_D$ elements at the first order, we generate a set of all unique three-orbital
    tuples by
    grouping all the 2-hole and 1-particle or 2-particle and 1-hole orbitals from the list of screened
    $T_2^{(1)}$. Thus the maximum number of tuple elements in the set is $4N_D$, but in practice 
    this number is significantly less. From Eq. \ref{second_order_wf}, note that any important rank-three operators
    that lead to three-body excited determinants must contain three orbital indices coming from 
    $T_2^{(1)}$, we thus target to retain only those rank-three operators that have at least 
    one of those unique three-orbital tuples. This significantly compactifies the search space for 
    the selection of the dominant $T_3$.
    
    \item \textbf{\textit{Excitation pathway pruning through leading order approximation of the scatterers:}} In the classical computer, to have the minimal computational overhead towards the generation and assessment of the important triple excitations, we retain only those $V_\mu$ in the summation in Eq. \eqref{T_3_compact} for which $|v_\mu/\Delta_\mu|>\epsilon_2$,
    where $\epsilon_2$ is another user defined threshold. It is important to note that the structure of $V_\mu$ is that of a two-body operator with effective hole-particle excitation rank of one. The contractible operators (either hole $(m)$ or particle type $(e)$) in Eq. \ref{second_order_wf} appear as a \textit{destruction} operator. As such one may introduce a \textit{scatterer operator} $S$ with identical hole-particle structure 
    as $V_\mu$ whose leading order measure will be $s_\mu^{(1)}=|v_\mu/\Delta_\mu|$. Here 
    $\Delta_\mu$ is a local 
    denominator that we had approximated to define the scatterer instead of the global denominator $\Delta_{\overline{K}}$ that appears in Eq. \eqref{T_3_compact}. The anti-hermitian operator
    $\sigma=\theta\hat{\kappa}$ will be used to denote $S-S^\dagger$. 
    \item \textbf{\textit{Excitation subspace pruning based on the overall contribution:}} Finally, with the two criteria above satisfied, we ensure that:
    \begin{equation}
     \Bigg| Amp\Bigg(\frac{(\contraction{}{V_\mu}{}{T_I^{(1)}} V_\mu T_I^{(1)})_{\overline{K}}}{\Delta_{\overline{K}}}\Bigg)\Bigg| > \epsilon_3  
    \label{selection_T3}
    \end{equation}
    Here \textit{Amp} suggests the amplitudes part of the corresponding equation.
    Even though the first two criteria above are satisfied, this criterion ensures that only 
    the most dominant resultant triples are finally included after cancellation of terms 
    with opposite signs. One must note that with the first two criteria (along with the 
    filtering of the two-body amplitudes at first order) above satisfied, the generation 
    of a term like $\contraction{}{V_\mu}{}{T_I^{(1)}} V_\mu T_I^{(1)}$ requires only a few 
    scalar multiplications and we will exploit this while choosing a dominant 
    unique pathway (\textit{vide infra}).
\end{itemize}
While one may explicitly include all the selected three-body cluster operators in the UCC ansatz, 
this leads to a pronounced escalation in the gate count and gate depth. Thus in order to 
keep the gate depth at its minimum, we decompose every selected $T_3$ as $T_{\overline{K}} \leftarrow (\contraction{}{S_\mu}{}{T_I^} S_\mu T_I)_{\overline{K}}$ as they appear in the second order 
expansion of the wavefunction Eq. \eqref{second_order}-\eqref{T_3_compact}. In the 
unitarized version, it is straightforward to see that every such contraction between
the operators can be captured by a commutator of the associated anti-hermitian operators:
\begin{equation}
    [\hat{\kappa}_\mu, \hat{\kappa}_I] \longrightarrow \hat{\kappa}_{\overline{K}}
    \label{sigmataucomm}
\end{equation}
Note that it is important to identify the appropriate operator label $\mu$ 
such that the scatterer $\hat{\kappa}_\mu$ does not commute with the excitation 
operators $\hat{\kappa}_I$. This means that the contractible
destruction operator in $S$ must be common to one of the creation orbital indices 
in $T$. To simulate such a high rank excitation, clearly one needs to pair up
appropriate $\tau$ and $\sigma$ in a factorized~\cite{Halder2023} manner such that
\begin{eqnarray}
& e^{\sigma}e^{\tau}=e^{{(\sigma+\tau)}+\frac{1}{2}[\sigma,\tau]+...}
\label{exppdt1}
\end{eqnarray}
At this stage it is important to note that for a 
\textit{canonical Hartree-Fock} reference, the triple excitation that appears in 
the leading order is represented by explicitly connected diagrams~\cite{windom2024new} as shown 
in Eq. \ref{T_3_compact}. Thus unlike the work by Fedorov \textit{et. al.}~\cite{Fedorov2022unitaryselective} who approximate the triples as disconnected 
product of two operators, our representation is manifestly connected.

\textbf{\textit{Selection of a unique pathway to triple and higher rank excitation manifold:}} While the decomposition of the high rank excitation in terms of lower rank operators 
is attractive, this may often lead to redundant generation of a given triple excitation at the 
cost of more number of parameters. 
There may be several pairs of $\hat{\kappa}_{\mu}$ and $\hat{\kappa}_{I}$ which leads to the 
same three-body excitation:
\begin{eqnarray}
    [\hat{\kappa}_{\mu_1}, \hat{\kappa}_{I_1}] \longrightarrow \hat{\kappa}_{\overline{K}} \nonumber 
\end{eqnarray}
\begin{eqnarray}
    [\hat{\kappa}_{\mu_2}, \hat{\kappa}_{I_2}] \longrightarrow \hat{\kappa}_{\overline{K}} \nonumber
    \text{\hspace{0.2cm} and so on...}
\end{eqnarray}
One may expand the implied sum in Eq. \eqref{selection_T3} into
sum of the associated scalar products between the selected $V_\mu$ and $T_I$:
\begin{eqnarray}
      Amp\Bigg(\frac{(\contraction{}{V_\mu}{}{T_I^{(1)}} V_\mu T_I^{(1)})_{\overline{K}}}{\Delta_{\overline{K}}}\Bigg) &=&  Amp\Bigg(\frac{ (V_{\mu_1}\boldsymbol{\cdot} T_{I_1}^{(1)})_{\overline{K}}}{\Delta_{\overline{K}}}\Bigg) \nonumber \\
      &+& Amp\Bigg(\frac{ (V_{\mu_2}\boldsymbol{\cdot} T_{I_2}^{(1)})_{\overline{K}}}{\Delta_{\overline{K}}}\Bigg)+ \hdots
      \nonumber\\
    \label{selection_T3_explicit}
\end{eqnarray}
and retain only that unique pair of ${\sigma}_{\mu}$ and ${\tau}_{I}$ at the second order for which 
the absolute value of individual scalar products appearing on the right side of Eq. \eqref{selection_T3_explicit} is maximum. Of course,
the corresponding $\tau_I$, which is already included at 
the first order but whose corresponding $[\sigma,\tau]$ 
pathway gets excluded at the second order through the 
unique pathway selection for triples, is not removed 
from the operator block since the corresponding doubles 
are important; we simply do not pair up any $\sigma$ along
with it.

With the knowledge of the first order corrected ansatz (Eq. \eqref{unitary1}) along with the 
important triples and the scatterers through the second order screening criteria 
stated above, we expand the various operator blocks in the ansatz 
with the second order estimate as:
\begin{equation}
    U^{[2]} = ...\Big[\overline{e^{\sigma_{\nu^{'}}}\hdots
    e^{\sigma_{\nu}}e^{\tau_L}}\Big]_{\delta}... \Big[e^{\tau_K}\Big]_{\gamma}... \Big[\overline{ e^{\sigma_\lambda}e^{\tau_J}}\Big]_{\beta}...
    \label{unitary2}    
\end{equation}
One may note that there may be more than one $\sigma$ that can couple with a given $\tau$ to
generate two or more distinct three-body terms, all of 
which may get selected as per the criteria stated before.
Thus one operator block may contain more than one 
$\sigma$ as shown above; however, there cannot be more than
one $\tau$ in any given operator block.
Furthermore, cases may arise where no $\sigma$ is chosen through the selection criteria 
to pair up with a $\tau$ to generate triples, as shown in the operator block $\gamma$ as an example.
In general,
\begin{equation}
    U^{[2]} = \prod_{\alpha=1}^{N_D} \Big[\overline{e^{\sigma_{\mu^{'}}}... e^{\sigma_{\mu}}e^{\tau_I}}\Big]_{\alpha}
    \label{unitary2gen}    
\end{equation}
where the overbars denote the implied commonality of the contractible index between the 
$\tau$ and $\sigma$. There may be fortuitous scenario 
due to the factorized structure of the unitary for which the arrangement of the operators 
simulates the third order effects of quadruple excitations. 
For example, in Eq. \ref{unitary2},
$e^{\sigma_{\nu^{'}}}$ may act on the composite $e^{\sigma_{\nu}}e^{\tau_L}$ that simulates a quadruple, whereas both $e^{\sigma_{\nu}}e^{\tau_L}$ and $e^{\sigma_{\nu^{'}}}e^{\tau_L}$ simulate two distinct triples. 

Along with the triples, the singles also are generated for 
the first time at the second order of MBPT. The selection 
procedure of the singles at the second order also closely
follows the way significant triples are selected. We choose
only those singles for which
    \begin{equation}
      \Bigg|Amp\Bigg(\frac{(\contraction{}{V_{\overline{\mu}}}{}{T_I^{(1)}} V_{\overline{\mu}} T_I^{(1)})_{\underbar{K}}}{\Delta_{\underbar{K}}}\Bigg)\Bigg| > \epsilon_3  
    \label{selection_T1}
    \end{equation}
Note that underbarred and capitalized letters, $\underbar{K}$ here represents the tuple of one hole 
and one particle index and $V_{\overline{\mu}}$ in Eq. \eqref{selection_T1} are two-body operators with
effective \textit{1p-1h} de-excitation structure. Unlike in Eq. \eqref{selection_T3}, no 
filtering for $V_\mu$ is employed here although the associated
$T_I^{(1)}$s are those which are previously selected at 
first order. Since $T_1$s are very few in number and 
contribute insignificantly to the gate depth, the selected
$T_1$s are parametrized as they are. Note that both in 
Eqs. \eqref{selection_T3} and \eqref{selection_T1}, we have usually employed the same threshold since both singles are triples
are first generated at the same order. Once the singles are screened, they are appended 
at the end of Eq. (\ref{unitary2gen}) and the final ansatz at the end of second order 
can be expressed as:
\begin{equation}
    U^{[2]} = \prod_{\beta=1}^{N_{S}}\Big[e^{\tau_{\underline{I}}}\Big]_\beta\prod_{\alpha=1}^{N_D} \Big[\overline{e^{\sigma_{\mu^{'}}}... e^{\sigma_{\mu}}e^{\tau_I}}\Big]_{\alpha}
    \label{unitary_with_singles}    
\end{equation}
where $N_{S}$ denotes the total number of singles screened.

\subsubsection{Higher order generalizations}
The connected quadruples are generated for the first time in the third order and
one may arbitrarily expand the ansatz by including leading 
order contributions by climbing up the MBPT ladder. Due to
the hierarchical structure of the MBPT equations, only the 
\textit{important} lower order terms may contribute to
higher order. Thus with the knowledge of the dominant triples 
and the dominant $\sigma_\nu$,
exactly the same protocol may be followed while 
moving to the third order generation of quadruply excited terms as one does
while moving from first to second order. The quadruples are thus 
simulated through the nested commutator: $[\hat{\kappa}_\nu,[\hat{\kappa}_\mu, \hat{\kappa}_{I}]] \longrightarrow \hat{\kappa}_{K=4}$ where $K=4$ 
denotes that eight orbital tuple index associated with a given quadruple excitation. 
Of course, additional care needs to 
be taken (if one desires so) to include only the unique quadruples that are not fortuitously included
at the second order due to the pairing up of more than one $\sigma$ to 
a $\tau$
in a given operator block as an artifact of the disentangled structure of the unitary.
At this stage, we mention that the third order is the 
lowest perturbative order where the unlinked diagrams appear and eventually 
get cancelled by the renormalization terms. Thus our approach towards the
construction of the dynamic ansatz satisfies linked cluster theorem and 
is perturbatively meaningful. 

The 
factorized structure of our ansatz (at any given 
perturbative order with various unitary
operators appearing as a product) bears a resemblance to 
the disentangled UCC theory (dUCC)~\cite{10.1063/1.5133059}. However, instead of 
ADAPT-VQE-like gradient based numerical selection and 
ordering of the operators from a pre-designed operator 
pool, we base our ansatz construction on perturbative 
analysis. While at the first order, it leads to the
conventional dUCCD ansatz with a specific operator ordering,
the higher order many-body perturbative structure naturally incorporate 
appropriate pair of scattering and cluster operators to be grouped 
in each operator block to induce high rank excitations via Eqs. 
\ref{sigmataucomm} and \ref{exppdt1}.

One may note that the various ranks of cluster
operators that are included or implicitly simulated are by construction
symmetry adapted, and we do not need to impose additional 
symmetry constraints. Furthermore, the entire ansatz construction pipeline does not
involve any measurement and is solely based on first-principle based perturbation theoretic 
measures. Due to no measurement overhead involved in COMPACT, the optimal structure of 
the ansatz is independent of any device noise. A bird's-eye-view of the 
operator selection scheme at various perturbative order is shown in Fig. \ref{fig:algorithm}. Below, we demonstrate the efficacy of the 
ansatz in capturing electronic strong correlation for various challenging cases.
\begin{figure*}
    \centering
    \hspace*{-1.8cm}\includegraphics[width=21.6cm, height=17.5cm]{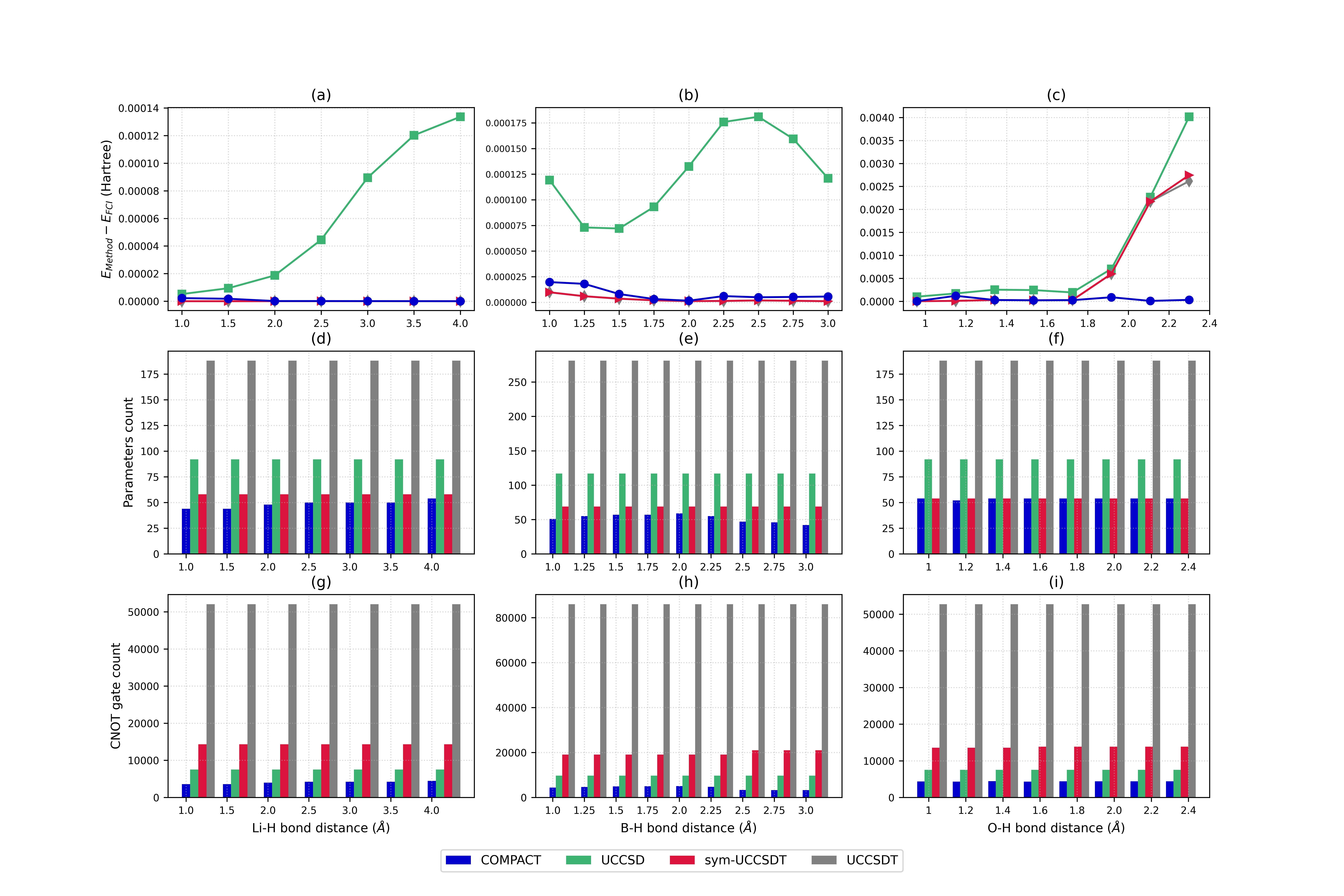}
    \caption{\textbf{Energy Error with respect to FCI (first row), parameter count (second row) and execution CNOT gates count per function evaluation (third row) over the potential energy profile for $LiH$ (first column), $BH$ (second column) and $H_{2}O$ (third column).}}
    \label{fig:pes_par}
\end{figure*}

\begin{figure*}
    \includegraphics[width=11cm, height=8.25cm]{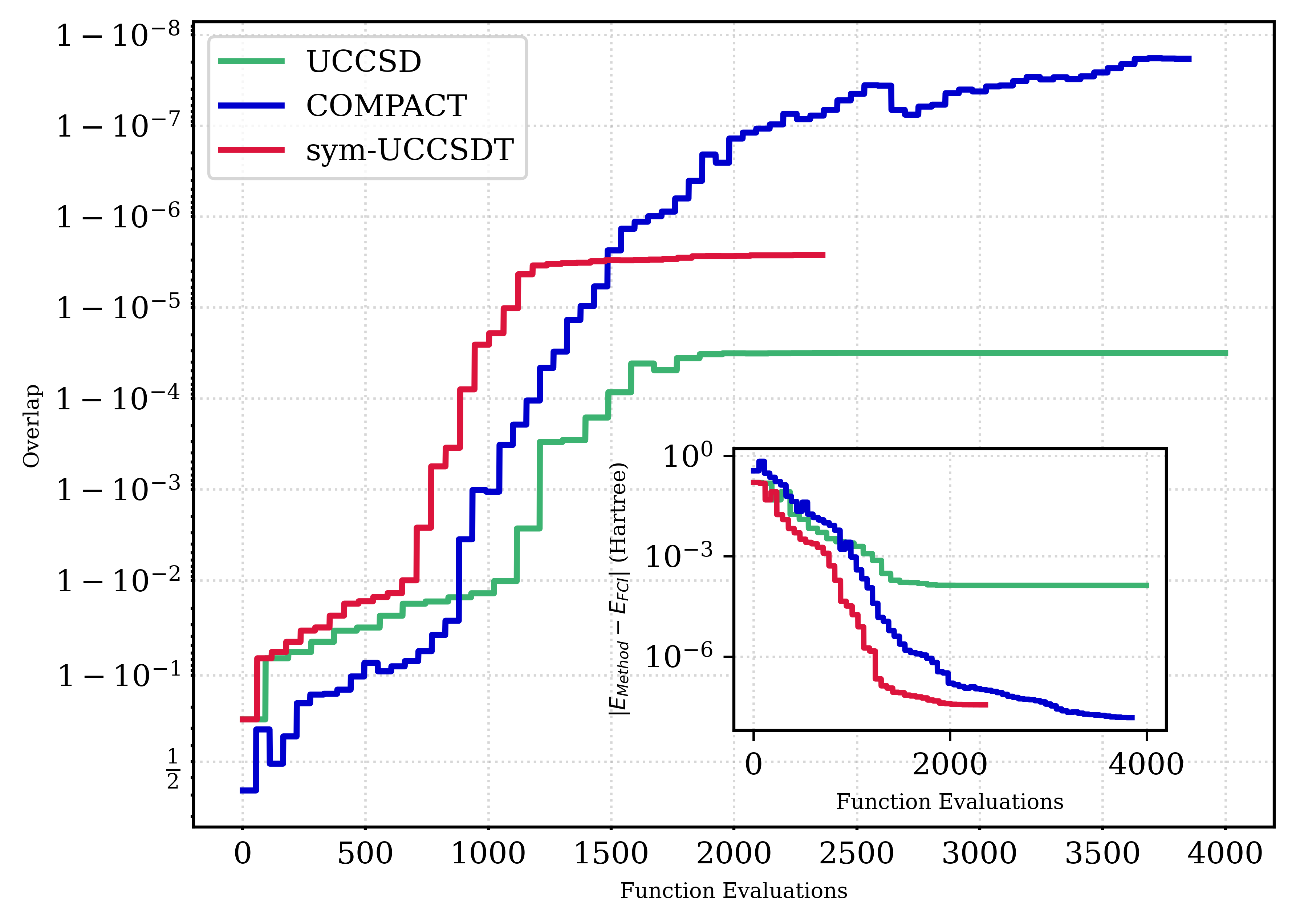}
    \caption{\textbf{Convergence trace as measured by the overlap with FCI wavefunction against the number of function evaluations for UCCSD, sym-UCCSDT and COMPACT for $LiH$ at bond length= 4\AA. The inset shows the convergence in predicted energies.}}
    \label{fig:convergence}
\end{figure*}

\section{Numerical results and discussions}\label{res}
\subsection{Methodology and general considerations}
In this section we illustrate the efficacy of our method by computing the ground state potential energy curves for three prototypical molecular systems: $LiH$, $BH$ and symmetric bond stretching of $H_{2}O$. For the numerical simulations, we have employed Qiskit nature~\cite{the_qiskit_nature_developers_and_contrib_2023_7828768} along with the PySCF module~\cite{https://doi.org/10.1002/wcms.1340}. For all the calculations we have employed the contracted STO-3G basis~\cite{doi:10.1063/1.1672392}. We have chosen the Jordan-Wigner~\cite{doi:10.1063/1.4768229} mapping technique to transform the second-quantized operators into the qubitized quantum circuit. For the optimization part of the VQE algorithm, we have worked with the gradient based L-BFGS-B~\cite{doi:10.1137/0916069} optimizer with the initial parameters for two-body operators set to their respective first order perturbative estimates and those corresponding to $T_{1}$ are set to zero. As the performance of COMPACT hinges upon the meticulous selection of the thresholds $\epsilon_{1}$, $\epsilon_{2}$ and $\epsilon_{3}$, we henceforth denote it as COMPACT(-log $\epsilon_{1}$, -log $\epsilon_{2}$, -log $\epsilon_{3}$). In all the cases, the reported parameter 
count accounts for the spin-complementary excitations as separate independent variables.

\subsection{Potential energy profile of strongly correlated systems: Accuracy, parameter count and CNOT gate count}
We demonstrate the efficacy and accuracy of COMPACT with three prototypical systems, namely 
the stretching of \textit{LiH, BH} and the simultaneous symmetric stretching of the 
bonds of $H_2O$, all of which exhibit intricate interplay of diverse electronic correlation 
effects. We first investigate the dissociation potential energy profile for $LiH$ molecule. Within 
the STO-3G basis set, the $LiH$ molecule is characterized as a 12 qubit (spinorbital) system spanned
by 4 electrons. In Fig. \ref{fig:pes_par}, we have plotted the energy errors with respect to the classically computed FCI energies (Fig. \ref{fig:pes_par}(a)), parameters count (Fig. \ref{fig:pes_par}(d)) and two qubit CNOT gates count (Fig. \ref{fig:pes_par}(g)) of our 
method for various $Li-H$ bond lengths ranging between $1\AA$ and $4\AA$, and compared them
against the conventional UCCSD, UCCSDT and sym-UCCSDT~\cite{doi:10.1021/acs.jpca.3c01753}. UCCSD with a parameter count (CNOT gate count) 
of 92 (7,520) shows a catastrophic failure for stretched bond lengths (see Fig. \ref{fig:pes_par}(a)),
hinting towards the poor expressibilty of the UCCSD ansatz in the regions of molecular strong 
correlation. UCCSDT, on the other hand, exhibits remarkable accuracy across the 
entire spectrum of the potential energy surface, however, at the expense of exceedingly large number 
of parameters (188) and two qubit CNOT gates (52,064). The sym-UCCSDT ansatz shows 
significant savings in quantum complexities without compromising the accuracies at the 
expense of only 58 parameters and 14,304 CNOT gates. Remarkably, the parameters count 
(CNOT gate count) corresponding to COMPACT(5,5,4) in the best case scenario 
(for geometries around the equilibrium) and the worst case scenario (for stretched geometries) 
are 44 (3,580) and 54 (4,480) respectively. It is worth pointing out at this juncture that although 
there is not significant difference between the sym-UCCSDT and COMPACT(5,5,4) in terms of 
parameter count, COMPACT shows significant reduction in the number of CNOT gates. With a very
conservative choice of thresholds, i.e., $\epsilon_{1},\epsilon_{2}=10^{-5}$ and 
$\epsilon_{3}=10^{-4}$, we get an order of magnitude reduction in gate count compared to 
sym-UCCSDT and UCCSDT. Such savings in circuit depth can be attributed to the fact that 
COMPACT encapsulates the effects of higher order excitations via only rank-two and rank-one 
excitations without explicitly involving the higher order excitations, unlike the other 
conventional methods. One may note that in the NISQ devices that suffer from various sources 
of detrimental noise, the reduction in the CNOT gate count (rather than the number of 
parameters) is more desirable, and COMPACT is a significant stride towards this.
Additionally for \textit{LiH} at a strongly correlated regime (bond length $4\AA$), we have 
plotted (Fig. \ref{fig:convergence}) the overlap of the final state (for UCCSD, 
sym-UCCSDT and COMPACT) with full configuration interaction (FCI) state and their energy 
convergence landscape against the number of function evaluations for all these methods. COMPACT
clearly demonstrates its superiority over other methods in predicting more accurate 
states and energy as compared to other methods under choice.\\

The next system under the consideration is $BH$; the potential energy profile when the $B-H$
bond is stretched is challenging to model due to interplay of strong and weak correlation and thus
this is considered as a standard testbed for evaluating the performance of a newly developed 
electronic structure methodology.
$BH$ in STO-3G basis can be identified as a 6 electron system distributed in 12 qubits 
(spinorbitals). In Fig. \ref{fig:pes_par}(b), we have plotted the energy deviations of COMPACT
and other allied conventional methods (UCCSD, UCCSDT, sym-UCCSDT) with respect to the FCI. 
While UCCSD exhibits an energy deviation $\sim 10^{-4}$ Hartree for $B-H$ distances beyond 
$1.75\AA$, the same for UCCSDT and sym-UCCSDT is of the order of $10^{-6}$ Hartree, clearly 
demonstrating the importance of higher excitations for strongly correlated regions.
With the fixed structure of UCCSD, UCCSDT and sym-UCCSDT, their associated parameters count 
(CNOT gates count) are 117 (9,672), 281 (86,020) and 69 (19,080 for $R_{B-H} < 2.5\AA$ and 20,976 for $R_{B-H} \ge 2.5\AA$) respectively. Note that for sym-UCCSDT in STO-3G basis, the change in the virtual orbital symmetries beyond $R_{B-H}=2.5\AA$ leads to the change in the number of CNOT count.
Interestingly, with thresholds $\epsilon_{1}, \epsilon_{2}=10^{-3}$ and $\epsilon_{3}=10^{-4}$,
COMPACT(3,3,4) follows an energy profile almost as good as UCCSDT and sym-UCCSDT. What is 
even more alluring, the quantum complexities \textit{i.e.}, parameters count (and the 
CNOT gates count) corresponding to the COMPACT(3,3,4) ansatz falls within a range: 59 (5,048) 
(at $R_{B-H}=2\AA$) 
and 42 (3232) (at $R_{B-H}=3\AA$), showcasing a tremendous reduction in quantum resources 
throughout the potential energy profile. The quantum complexities associated with 
COMPACT and the other conventional 
ans\"{a}tze has been illustrated along the various $B-H$ bond lengths in 
Fig. \ref{fig:pes_par}(e) and \ref{fig:pes_par}(h).

The final system under consideration is the symmetric simultaneous stretching of $O-H$ bonds
in $H_{2}O$ molecule. With the 1s core orbital of $O$ frozen, $H_{2}O$ in STO-3G basis renders
to be a 8 electrons system in 12 spinorbitals. Fig. \ref{fig:pes_par}(c) shows the energy accuracy 
over the potential energy profile ($O-H$ bond lengths ranging between $R_{e}$ and $2.4$ times $R_{e}$ with $R_{e}=0.958\AA$ and 
the $H-O-H$ angle kept fixed at 104.4776$\degree$) compared to FCI. Fig. \ref{fig:pes_par}(f) 
and \ref{fig:pes_par}(i) showcases a comparative study of the corresponding quantum complexities 
(parameters count and two qubit CNOT gate count) associated with COMPACT, UCCSD, UCCSDT and sym-UCCSDT. 
While UCCSD requires 92 (7552) parameters (CNOT gates), it fails catastrophically with the 
stretching of $O-H$ bonds when the molecular strong correlation kicks in. Unexpectedly, both 
UCCSDT and sym-UCCSDT too shows similar trend as UCCSD. A possible explanation for such a behaviour 
might be the inability of both the ans\"{a}tze in navigating towards the absolute minima. On the other 
hand, COMPACT (5,5,4) shows remarkable accuracy throughout the potential energy profile at the 
expense of mere 52-54 parameters and 4308-4440 CNOT gate counts. One may note that the quantum resources
deployed by COMPACT is at least one order of magnitude less than those required even by 
sym-UCCSDT; however, 
the former ascertainably outperforms the conventional ans\"{a}tze in terms of accuracy throughout the 
energy profile. We reiterate that such dramatic reduction in quantum complexity, particularly in
the CNOT gate count, makes the realization of COMPACT in NISQ devices seemingly feasible. 

Towards the end of this section, we also highlight the scalability of 
this approach to treat realistic many-body systems. The COMPACT ansatz pre-screen the 
many-body bases and retains only the ``most dominant" terms of the \textit{N}-electron 
Hilbert space \textit{at each perturbative order}. These excitations are 
subsequently included in the construction of COMPACT ansatz via lower rank 
decomposition in terms of the cluster operators 
and a cascade of scatterers. The hierarchical selection of the dominant determinants 
and the screening at each perturbative order to retain their unique pathway 
further ensure to restrict the number of 
parameters and CNOT gate count to their minimum. With the increase in system size, 
this does not approach the exponential quantum complexity of Full Configuration
Interaction as the number of such determinants screened by COMPACT 
at any given perturbative order practically increases sub-exponentially with the 
system size.

\section{Conclusions and future outlook}\label{concl}
In this study, we have proposed a novel algorithm towards an intuitive construction of a dynamic 
structured ansatz in a system-specific manner. COMPACT caters to the dual objective of minimization 
of circuit depth while concurrently eliminating the necessity of any pre-circuit measurements 
towards the construction of an optimally expressive ansatz. Unlike most of the prevalent dynamic
ans\"{a}tze, COMPACT relies on classical approximations for ansatz construction and compactification; 
thus, it is safeguarded from any kind of quantum hardware induced noise. Built upon the many-body
perturbation theory, the accuracy and quantum complexities associated with COMPACT are reliant only 
on the perturbative order and hence, are tunable.\\
Through a number of numerical applications on prototypical strongly correlated molecular systems, 
we have demonstrated that the resulting ansatz is expressive enough to handle molecular strong
correlation with only a shallow circuit depth and zero measurement overhead in its construction process.
While the genesis of the ansatz is performed ``classically", it still 
requires quantum measurements for the iterative optimization of the parameters and thus
the final outcome (but not the structure of the ansatz) is likely to get impacted by hardware noise. 
In this regard, 
it would be interesting to introduce CNOT-efficient
circuits~\cite{PhysRevA.102.062612,Yordanov2021, doi:10.1021/acs.jctc.2c01016} with additional error mitigation routines~\cite{10.1063/5.0166433} 
for COMPACT which is likely to be further resilient to hardware noise during its execution. COMPACT may also be implemented in terms of the Quantum 
Flow algorithm~\cite{PhysRevLett.131.200601} in reduced dimensionality 
active spaces to further minimise its final circuit complexity.
An excited state extension to COMPACT will also be highly desirable and will be a subject 
of our future endeavors in this direction.

\section*{Acknowledgements}
D.H. thanks the Industrial Research and Consultancy Center (IRCC), IIT Bombay and D.M. acknowledges the Prime Minister’s Research Fellowship (PMRF), Government of India for their research fellowships. 
\section*{Conflict of Interests}
The authors have no conflict of interests to disclose.
\section*{Data Availability}
The data that support the findings of this study are available from the corresponding author upon reasonable request.


%


\end{document}